# IMPACT OF INDENTATION IN PROGRAMMING

Parvatham Niranjan Kumar[1]

[1]Department of Information Technology, Anurag Engineering College,Kodad,Andhra Pradesh,India

*ABSTRACT*

*In computer programming languages, indentation formats program source code to improve readability. Programming languages make use of indentation to define program structure .Programmers use indentation to understand the structure of their programs to human readers. Especially, indentation is the better way to represent the relationship between control flow constructs such as selection statements or loops and code contained within and outside them. This paper describes about different indentation styles used in Programming and also describes context of each indentation style. It also describes indentation styles used for various programming constructs and the best practice for a particular programming construct. This paper helps the beginners to understand various indentation styles used in programming and also to choose suitable indentation style.*

## 1. INTRODUCTION

In programming languages, indentation formats program source code to improve readability. Programming languages make use of indentation to define structure of the program.

### 1.1. IMPORTANCE OF INDENTATION IN PROGRAMMING

The following are some of the reasons to maintain indentation in programming
- It shows levels of nesting.
- Anyone reading your code can tell what's executing inside of what
- It shows scope
- Easier to read
- Better programming sense

### 1.2. SIZE OF INDENT

The size of the indent is independent of the style that is used to write program. In Most of the early programs tab characters are used to format source code, by that we can save size of the file. In Linux/UNIX editors' a tab is equivalent to eight characters, while In the case of Macintosh or Microsoft Windows environment a tab is set to four characters, this creates confusion when code was transferred from one environment to another. But Modern editors are able to choose indentation sizes randomly, and will insert the appropriate combination of spaces and tabs. Spaces instead of tabs increase cross-platform functionality.

## 2. STYLES OF INDENTATION

The following section describes various indentation styles for good programming. Each style has its own context to use.

### 2.1. K & R Style

In this style every function consists of opening brace in the next line at the same indentation level as the header of the function, the code within the braces is indented, and the closing brace at the end is on the same indentation level as its header. It is commonly used in C, C++, and C #. The following code describes this style.

```
int main(int argc, char *argv[])
{
   ...
   while (x == y) {
      some();
      some_else();

      if (some_error) {
         do_correct();
      } else
         continue_as_usual();
   }
}
```

### 2.2. Variant: 1TBS Style

In 1TBS style, constructs keeps opening brace on the same line and code lines are on separate lines, and constructs that do not allow insertion of new statements are on a single line . The source code of the UNIX kernel was written in this style. The following code fragment describes about this style. This style reduces number lines of code because beginning brace does not require an extra line.

```
if (x < 0) {
       puts("Negative");
       negative(x);
   } else {
       puts("Non-negative");
       nonnegative(x);
   }
```

### 2.3. Variant : Stroustrup Style

This is Bjarne Stroustrup's version of K&R style for C++,Stroustrup enhances K&R style for classes . This style keeps opening brace of class, at the same line as name of the class .Stroustrup style is suitable for short functions by keeping all on one line. Stroustrup style used as a indentation style in the editor Emacs. The following code is the example for this style.

```
class Sample {
   public:
      Sample(int s) :elemt{new double[s]}, sz{s} { }
      double& op[](int i) { return elemt[i]; }
   };
```

**2.4. Allman Style**

Eric Allman invented this kind of style. In this the opening brace associated with a control statement at the next line, indented to same column as the first letter in the name of control statement. Code within the opening brace and closing brace is indented to the next level. In this style the indented code is set apart from the containing statements to improve readability, and the closing brace indented to the same column as the opening brace, it makes easy to find matching braces. The following code describes about this style.

```
while (x == y)
{
   Do_something();
   Do_somethingelse();
}
```

**2.5. BSD KNF Style**

It is also called as Kernel Normal Form, this kind of style used for the code used in the Berkeley Software Distribution operating systems. The hard tabulator is kept at 8 columns, while a soft tabulator is often defined as a helper and set at 4.The following code describes about this style.

```
while (x == y) {
       Do_something();
       Do_somethingelse();
}
```

**2.6 Whitesmiths Style**

This style keeps the brace associated with a control statement on the next line, and indented to the next level. Code within the braces is lines up to the same level as the braces. It is used in the documentation for the first commercial C compiler. The advantage of this method is that the alignment of the braces with the block emphasizes the fact that the entire block is conceptually a single compound statement. Indenting opening and closing braces gives clear idea that they are subordinate to the control statement. The problem with this style is that the ending brace no longer lines up with the control statement it belongs to. The following code fragment describes Whitesmiths style.

```
while (x == y)
   {
   Do_something();
   Do_somethingelse();
   }
```

**2.7 GNU Style**

This style keeps braces on a separate line, indented by 2 spaces, except the opening and closing brace function, where they are not indented. In other cases, the contained code is indented by 2 spaces from the braces. This style adds the benefits of Allman and Whitesmiths, and removes the difficulties in Whitesmiths style. The following code fragment describes about this style.

```
combine (char *str1, char *str2)
{
  while (x == y)
```

```
    {
     Do_something ();
     Do_somethingelse ();
    }
  Do_final ();
}
```

### 2.8 Horstmann Style

This style uses Allman style by keeping the first statement of a code block on the same line as the opening brace. This style uses the concept of Allman by keeping the vertical alignment of the braces to improve readability and easy to identify of code blocks. The following code fragment describes this style.

```
if (x == y)
{   Do_something();
    Do_somethingelse();
    //...
    if (y< 0)
    {   printf("Negative");
        negative(y);
    }
    else
    {   printf("Non-negative");
nonnegative(x);
    }
}
```

### 2.9. Pico Style

The style normally used in the Pico programming language In this style the beginning and closing braces are both share space with a line of code. The following code describes about pico style.
```
stuff(n):
{ x: 3 * n;
  y: doStuff(x);
  y + x }
```

### 2.10. Banner Style

This style makes visual scanning easy for reader. Headers" of blocks are the only thing indented at that level. The following block of code describes about this style.

```
Sample1 () {
  do something
  do somemorething
  }
 Sample2 () {
  etc
  }
```

## 2.11. Lisp style

In Lisp style programmer can insert closing brackets in the last line of a code block. Indentation is the only way to identify blocks of code. This style has the does not keep uninformative lines. The following code describes about this style.

```
for (i = 0; i < 10; i++) {
   if (i % 2 == 0) {
      doSomething (i); }
   else {
      doSomethingElse (i); } }
```

## 2.12. Ratliff style

This style begins much like 1TBS but then the closing brace lines up with the indentation of the nested block. The following code fragment describes this style.

```
for (i = 0; i < 10; i++) {
   if (i % 2 == 0) {
      doSomething(i);
      }
   else {
      doSomethingElse(i);
      }
   }
```

## 3. INDENTATION FOR PROGRAMMING CONSTRUCTS
## 3.1. Functions

To indent the function, just apply the "Golden Rule" -. Always indent the statement of a code block with uniform amount from the first character of the control statement. Apply the same rule to the variable declaration types, too, but it is preferred to take step further and place each variable on a separate line. The following main function describes how to indent functions.

```
int main (int argc, char *argv[])
{
int     a,      /* an integer variable */
b;      /* another lousy comment */
float   c;      /* always write informative comments! */

printf("Give me an 'a' : ");
scanf("%d",&a);
printf(" Give me a 'b' : ");
scanf("%d",&b);
. . .
}
```

Function with a parameter list is longer than the length of the line, then divide the parameter list across lines, and indent the second line and subsequent lines with a sufficient amount so that the reader can able to understand parameter list apart from rest of the code in the function. The following function "Index_search" is an example for function with parameter.

```
int Index_search (struct collection    list[],
                  int                  value_first_index,
                  int                  value_last_index,
                  key_type             value_target         );
```

## 3.2. Selection Statements: IF, IF-ELSE, and SWITCH
### 3.2.1. IF with a Simple Body

If the body of an IF statement contains only one statement, then no need keeping set of braces. If the statement is short enough, it is better to keep it on the same line. If you place body on the next line, then it is acceptable to place braces around it. Later if you add a statement to the body, no need to remember adding braces to the compound statement. The following statement is an example for indentation IF with simple body.

```
if (count < 0) count = 0;
```

### 3.2.2. IF with a Compound Body

Statements of the Compound body are indented uniformly and consistent with chosen brace-placement style.

### 3.2.3. IF-ELSE with Simple Bodies

The same indentation used as the ordinary IF statement. Braces are not necessary. If braces are added  stick with  adopted brace-placement style. The following explains about this style.

```
if (temperature < 55)
        printf("It could be warmer...\n");
else
        printf("It could be colder...\n");
```

### 3.2.4. IF-ELSE with Compound Bodies

In this case the braces are required. Just be consistent with particular style. The following code fragment is an example for this style.

```
if (victory(wh))
   {
        h_wins++;
        printf("Human wins");
   }
 else
  {
        c_wins++;
        printf("Computer wins\n");
  }
```

### 3.2.5. IF-ELSE with Mixed Simple and Compound Bodies

This case becomes uncomfortable, when you leave off the braces for simple bodies so it is better practice to keep braces even for simple bodies. The following is the example code fragment for this style.

```
    if (sal > 2500) {
            tax = sal * 0.025;
            printf("Do you need a loan?\n");
} else {
            tax = 0;
}
```

### 3.2.6. Switch

The regular method to apply indentation for SWITCH is to align the CASEs with the SWITCH statement, and then indent the statements within the case from the level of the CASE statement. It is an extension to the IF-ELSE indentation scheme. But the problem with this approach is it is very hard to recognize the end of the SWITCH. So the better indentation schema is, indent the CASEs from the SWITCH, and then indent the statements from the CASES.

```
switch(ch) {
        case ' ': printf("space,\n");
                    break;
        case 'a':
        case 'e':
        case 'i':
        case 'o':
        case 'u': printf("a vowel.\n");
                    break;
        default : printf(" something else.\n");
}
```

## 3.3. Iteration Statements

It is easy to handle loops, if you have an idea of indentation to IF and IF-ELSE.

### 3.3.1. WHILE with a Simple Body

If the statements in the body of the WHILE loop is short, It is better to place on the same line as the loop statement. Instead, it is better habit to place the indented body on the next line with braces.

```
while ((i < LIST_SIZE) && (list[i] != MAX_SCORE)) {
        i++;
}
```

### 3.3.2. WHILE with a Compound Body

Statements of the Compound body are indented uniformly and consistent with chosen brace-placement style.

```
while (i < length) {
        printf("Element id and element %d is %d.\n",i,list[i]);
        i++;
```

### 3.3.3. FOR with Simple Body

It is better to keep body on the next line. Otherwise it makes reader little bit confusing.

```
for (i=0; (i < LIST_SIZE) && (list[i] != MAX_SCORE); i++);
        if (i >= LIST_SIZE)
        printf("Sorry; %d was not found.\n",MAX_SCORE);
```

In the above code there is no body for the loop. But the reader may believe that "if" statement is the loop's body.

The better way is use comments to specify null loop body.

```
for (i=0; (i < Lenght) && (list[i] != M_SCORE); i++)
        /* null body */ ;
if (i >= Length)
        printf("Sorry; %d was not found.\n",M_SCORE);
```

### 3.3.4. FOR with a Compound Body

In this case braces are required to keep the body statements. The following code is an example.
```
for (sum=0, i=0; i < LIST_SIZE; i++)
{
        printf("Element %d is %d.\n",i,list[i]);
        sum += list[i];
}
```

### 3.3.5. DO-WHILE with a Simple Body

If the DO-WHILE contains a simple body, then keep the whole loop on one line. The following code is an example for this style.
```
do i++;  while ((i < LIST_SIZE) && (list[i] != MAX_SCORE));
```

the best practice is always add braces around loop bodies,
```
i = 0;
do {
        i++;
} while ((i < LIST_SIZE) && (list[i] != MAX_SCORE));
```

### 3.3.6. DO-WHILE with a Compound Body

In this case, you need braces around the body.
```
i = 0;
do
{
        printf("Element %d is %d.\n",i,list[i]);
        i++;
} while (i < LIST_SIZE);
```

We can save one line by keeping "do" and "{"on the same line.
```
i = 0;
do {
        printf("Element %d is %d.\n",i,list[i]);
        i++;
} while (i < LIST_SIZE);
```

### 3.4. Nested Statements

Indentation is necessary if the code contains complex statements as part of other complex statements. Without indentation the reader may believe that a statement is outside of the body. Specifying "Levels" of Indentation for lines in the code makes reader easy to understand "level" of indentation applied to that line. The below code fragment depicts indentation of "Nested Statements".

```
[0] int main (void)
[0] {
[1]     int    k;
[1]     float  sum = 0, thousandth = 0.001;

[1]     for (k=1; k<=1000; k++) {
[2]         sum += thousandth;
[2]         if (k% 100 == 0)
[3]             printf("After %4d iterations, sum = %.10f\n",k,sum);
[1]     }
```

## 4. CONCLUSIONS

Good style programming is about making a program clear and understandable as well as easily modifiable. Now a day's students those who are learning programming are not giving much importance to the indentation to produce a code which is easily understood by the reader. They are not realizing with the importance of indentation and how it will impact on programming. This paper attempted to help beginners to understand about indentation styles used in programming.

### ACKNOWLEDGEMENTS

I would like to thank everyone, who encouraged me during the progress of this work.

### REFERENCES

[1]  "Tabs versus Spaces:" An Eternal Holy War. by Jamie Zawinski 2000

[2]  J. Lions (June 1977). "Unix Operating System Source Code Level Six". University of New South Wales.

[3]  Reddy, Achut (2000-03-30). "Java Coding Style Guide". Sun Microsystems. Retrieved 2008-05-30.

     Java Code Conventions". Sun Microsystems. 1997-09-12. Retrieved 2008-05-30.

[5]  Code Conventions for the Java Programming Language". Sun Microsystems. 1997-03-20. Retrieved 2008-05-30.

[6]  Bjarne Stroustrup (September 2010). "PPP Style Guide".

[7]  "Formatting Your Source Code". GNU Coding Standards.

[8]     Lammers, Susan (1986). Programmers at Work. Microsoft Press. ISBN 0-914845-71-3.

[9]     Linda Lamb, Learning the vi editor. O'Reilly

[10]    www.cs.arizona.edu

[11]    http://www.cprogramming.com

[12]    http://www.cs.armstrong.edu

**P.Niranjan Kumar** Received B.Tech degree in Information Technology from Jawahar Lal Nehru Technological University, Hyderabad,India and M.Tech in Computer Science Engineering from Jawaharlal Lal Nehru Technological University ,Hyderabad,India.Research interests include programming languages, Distributed omputing, 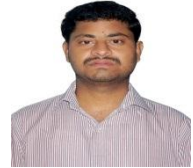